# Growth and Spectroscopic Properties of Pr$^{3+}$ Doped Lu$_2$S$_3$ Single Crystals

Vojtěch Vaněček,* Vítězslav Jarý, Robert Král, Lubomír Havlák, Aleš Vlk, Romana Kučerková, Petr Průša, Jan Bárta, and Martin Nikl

**ABSTRACT:** For the first time, Lu$_2$S$_3$ (undoped and Pr-doped) single crystals were successfully grown from a melt using a micropulling-down (mPD) technique. Customization of the halide mPD apparatus allowed us to grow rod-shaped (Ø 2 mm and a length of around 20 mm) crystals of Lu$_2$S$_3$ with a high melting temperature (∼1750 °C). X-ray powder diffraction revealed that the grown crystals exhibit an $\varepsilon$-Lu$_2$S$_3$ crystal structure ($\alpha$-Al$_2$O$_3$ type, space group $R\bar{3}c$). Optical and scintillation properties of both undoped and Pr$^{3+}$-doped Lu$_2$S$_3$ were investigated. Fast 5d→4f Pr$^{3+}$ luminescence was observed in both photoluminescence and radioluminescence spectra. The presented technology is an effective tool for the exploration of a large family of high melting sulfides. Such materials show promise for applications as scintillators, active laser media, and optoelectronic components.

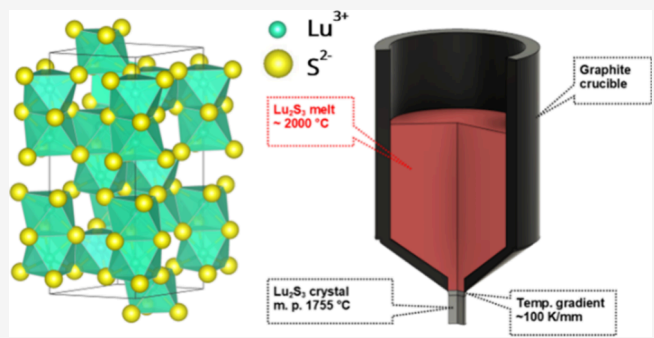

## INTRODUCTION

Rare-earth (RE) binary sulfides of general formula {RE}$_2$S$_3$ (where RE = Sc, Y, and La−Lu), also commonly called sesquisulfides, possess several attractive properties such as a high refractive index, a wide transparent region, and a wide region of homogeneity of physical properties.[1] However, their structural properties are very complex, as rare-earth sulfides exhibit polymorphism with more than seven different crystal structures.[2] The most common crystal structures reported for RE sesquisulfides with their usual empirical notation are given in Table 1. In addition to the listed sesquisulfide structures, RE sulfides with different stoichiometry, i.e., {RE}S, {RE}$_3$S$_4$, and {RE}S$_2$, some of which are not line compounds, but have rather wide homogeneity windows. P−T phase diagrams of both Yb$_2$S$_3$ and Lu$_2$S$_3$ were presented for the first time by Kanazawa et al.[3] RE sesquisulfides have very good chemical and mechanical stability. Most of the compounds melt above 1600 °C,[4] and their chemical resistivity is comparable to that of zinc sulfide as well as that of indium sulfide.

Due to the high melting points of RE sesquisulfides (above 1490 °C), difficulties are experienced when growing their crystals from melts of stoichiometric composition. To prevent decomposition, crystallization of rare-earth sesquisulfides from the melt in an atmosphere of sulfur vapor was suggested.[7] Selected rare-earth sesquisulfides were grown already in the 1960s (Dy, Gd, Nd, and Y) by either flame fusion (argon plasma torch) or the Bridgman technique.[8] Their melting points vary from 1490 °C for Dy$_2$S$_3$ to 2080 °C for La$_2$S$_3$.[8] So far, most of the attempts to grow single crystals of these binary sulfides have been limited to the production of small-sized crystals with less than 1 mm$^3$ in volume. Very often, chemical transport was used, see for example ref 9, where the starting sulfide powder was sealed in an evacuated quartz ampule together with iodine flakes (99.998% pure, concentration of 10 mg cm$^{−3}$). To achieve suitable transport rates, the temperatures were set to 1200 °C at the source and 1150 °C at the crystallization zone of the furnace. The growth procedure was always limited to a few days to avoid serious degradation of the quartz glass. The growth of Gd$_2$S$_3$ yielded prismatic needle-like crystals with maximum dimensions of 1.0 × 1.5 × 5 mm$^3$. X-ray diffraction confirmed the presence of the orthorhombic $\alpha$-Gd$_2$S$_3$ phase (see Prewitt and Sleight[10]). La$_2$S$_3$ yielded compact yellow crystals with many faces of the $\beta$-type tetragonal structure.[5] Y$_2$S$_3$ grew as yellow prismatic needles with a monoclinic structure of the $\delta$-type phase.[11] Similarly, the single crystals of $\alpha$-Sm$_2$S$_3$ and $\alpha$-Dy$_2$S$_3$ were grown by a chemical transport reaction method with iodine as a carrier gas.[12,13] The growth resulted in needle-shaped crystals with dimensions of a few millimeters along the $b$-axis. In the six-sided cross section (in the $ac$-plane), the dimensions were submillimeter, resulting in an overall weight of several

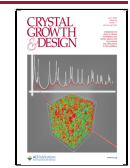

4736

Table 1. Most Common Crystal Structures Reported for RE Sesquisulfides[b]

| common empirical notation | α | β | γ[a] | δ | ε | ζ |
|---|---|---|---|---|---|---|
| crystal system | orthorhombic | tetragonal | cubic | monoclinic | rhombohedral | orthorhombic |
| space group | $Pnma$ | $I4_1/acd$ | $I\bar{4}3d$ | $P2_1/m$ | $R\bar{3}c$ | $Fddd$ |
| space group no. | 62 | 142 | 220 | 11 | 167 | 70 |
| phase prototype | $Cr_3C_2$ | $Pr_{10}S_{14}O$ | $Th_3P_4$ | $Ho_2S_3$ | $\alpha\text{-}Al_2O_3$ | $Sc_2S_3$ |

[a]{RE}-deficient phase derived from a {RE}$_3$S$_4$ structure. [b]For more information, see refs 2, 3, 5, and 6 and references herein.

milligrams. A gold-colored needle (dimensions, 0.10 mm × 0.03 mm × 0.4 mm) of $Er_2S_3$ was also grown by the vapor transport technique by Fang et al.[6] Samples prepared by quenching from high pressures and temperatures revealed a new orthorhombic ($Pnma$) modification of $Ln_2S_3$ (Ln = Y and Ho−Lu) with the $U_2S_3$ structure.[14]

Considering the melting point of $\varepsilon$-$Lu_2S_3$ of 1755 °C, the heaviest out of {RE}$_2$S$_3$,[4] repeatable growth of the $Lu_2S_3$ bulk single crystals should be viable using advanced growing techniques, such as micropulling-down (mPD) or Bridgman techniques. However, studies dealing with the optical and scintillation properties of $Ln_2S_3$ are very rare, especially for bulk single crystals. Therefore, there is significant potential for new and original research. Although $Ce^{3+}$-doped crystals of $Lu_2S_3$ have already been grown by chemical vapor transport as a potentially interesting scintillator with high density (6.25 g/cm$^3$), effective atomic number (66.7), and light yield approaching 30000 ph/MeV,[15] only very small crystals were obtained (0.2 × 0.4 × 2 mm$^3$), which were not suitable for applications. Moreover, contamination of the prepared crystals by iodine cannot be excluded. A content of 0.5% mol $Ce^{3+}$ in the $Lu_2S_3$ crystal was estimated based on the assumption that the absorption peak cross section for $Ce^{3+}$ in $Lu_2S_3$ is equal to that in $\beta$-$La_2S_3$:Ce.[15] There has been no systematic research investigating the optical properties of such compounds. The micropulling-down technique can be used for crystal growth with an accurate dopant content based on the exact composition of the input raw materials.

Furthermore, Cunningham et al.[16] showed that $Lu_2S_3$ with a corundum structure (space group $R\bar{3}c$) undergoes a reversible structural transition to the $Th_3P_4$-type structure (at ∼50 kbar) and that the band structure changes from a direct band gap ($E_g$ ∼ 3.2 eV at ambient pressure) to an indirect band gap (estimated $E_g$ = 2.0−2.5 eV at ∼100 kbar) as a result of the pressure-induced structural transformation. Typical 5d→4f emission of the $Ce^{3+}$ in $Lu_2S_3$ was shifted from 580 nm at ambient pressure down to 720 nm at 98 kbar.[16] Single-crystal samples doped with Nd were grown using a chemical vapor transport method,[9] and $Nd^{3+}$ spectroscopic properties in near-infrared regions were established.

In this work, we focus on the preparation of the bulk $Lu_2S_3$ single crystals (undoped and $Pr^{3+}$-doped) by an mPD technique using a customized halide mPD apparatus. The aim is to investigate the structural, optical, and luminescence properties of the $Lu_2S_3$ crystals. Obtained results are presented for the very first time.

## ■ EXPERIMENTAL SECTION

**Sample Preparation and Crystal Growth.** As a first step, sulfide powder samples were synthesized as precursors for subsequent crystal growth. Undoped $Lu_2S_3$ and Pr-doped $Lu_2S_3$ powders were prepared from powder $Lu_2O_3$ (5 N purity) and $Pr_6O_{11}$ (5 N purity) oxides, respectively, under the flow of $H_2S$ (2.5N purity) at a temperature of 1450 °C for 6 h in a graphite boat in an alumina tube furnace. The heating rate was 7 °C/min up to 700 °C, during which the alumina tube was continuously flushed with 5 N argon gas (30 L/h). From 700 to 1450 °C, $H_2S$ (1.5 L/h) was applied during the rest of the synthesis, including the cooling down to 250 °C. Then, argon (5 N) was applied again for flushing the furnace and during the cool-down to room temperature (RT). The remaining $H_2S$ flowing out of the furnace was absorbed in a NaOH solution. The resulting products ($Lu_2S_3$ and $Lu_2S_3$:Pr) were in the form of gray, sintered blocks. Afterward, the product was ground to a powder using a mortar and pestle. X-ray powder diffraction proved that all products were phase-pure $Lu_2S_3$. It is worth mentioning that $Lu_2S_3$ is not hygroscopic and was observed to be stable when exposed to the ambient atmosphere.

Crystals of lutetium sulfide, undoped and doped with praseodymium ($Lu_2S_3$:Pr), were grown by the mPD technique using a customized halide mPD apparatus (further denoted as H-mPD). The customization allowed the growth of rod-shaped crystals with 2 mm in diameter and a length of around 20 mm. The H-mPD setup is designed for crystal growth in a closed system under an inert atmosphere (usually 6 N Ar) to prevent degradation of the grown material via a reaction with oxygen or moisture. Several customizations had to be made for the successful growth of high melting $Lu_2S_3$ (mp = 1755 °C[4]) in the H-mPD apparatus, which is designed for the growth of low melting halides (mp < 1000 °C). Three layers of shielding made of fused alumina (including lids) were used to suppress heat losses. A reflector made of a platinum sheet was placed under the seed to decrease the temperature gradient below the crucible and to further suppress the heat losses. A custom-made water-cooling head was mounted on top of the growth chamber to prevent damage due to the difference in the thermal expansion coefficients of used construction materials. During the typical growth, the growth chamber was assembled in a glovebox (MBraun LABstar) under a dry $N_2$ atmosphere (concentration of $O_2$ and $H_2O$ below 1 ppm). The hot zone (graphite crucible, after heater, and 3 layers of alumina shielding with lids) was placed into a vacuum-tight growth chamber and enclosed with a fused silica tube (for details, see ref 17). The fused silica tube allows for in situ observation of the solid−liquid interface using a CCD camera. After the assembly, the growth chamber was closed with a vacuum valve and transported to the H-mPD apparatus. Before growth the chamber was connected to the rest of the apparatus, the apparatus was evacuated with a rotary vane pump (to ca. 1 Pa) and refilled with 6 N argon three times. The whole system was then degassed by heating the hot zone to ∼200 °C under a high vacuum ($10^{-4}$ Pa) using a turbomolecular pump. Afterward, the growth chamber was filled with 6 N Ar, and the growth proceeded in a very similar way to a typical H-mPD experiment. The basic steps of such growth include melting of the $Lu_2S_3$ powder, seed touch, growth (crystallization of the melt induced by downward movement of the seed through a temperature gradient), and cooling to room temperature. The heating and cooling rates were 20 and 15 K/min, respectively, and a typical growth rate was 0.1 mm/min. During high-temperature growth, graphite dust was released from the hot zone elements. This resulted in the presence of graphite both on the surface and inside the grown crystal. Under the growth conditions described above, the $Lu_2S_3$ melt is not wetting the crucible. In such a case, the maximum diameter of the mPD grown crystal is given by the inner diameter of the nozzle at the bottom of the crucible (for more information, see Fukuda and Chani[18]).

The as-grown crystals were cut and polished to obtain samples for optical characterization. A stereomicroscope photo of a disc prepared from $Lu_2S_3$:Pr (0.05%) shows that the sample is transparent (see Figure 1). However, the central section contains cracks around a black inclusion. The black inclusion is most probably graphite that was released from the hot zone elements during the growth. Further inspection of the sample using cross-polarizers showed that the uncracked section consists of two main grains.

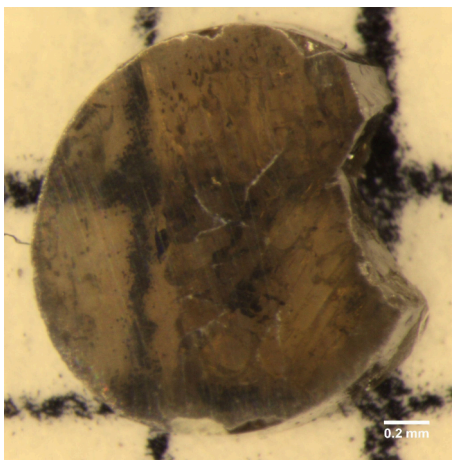

**Figure 1.** Picture of the cut and polished disc from the Lu$_2$S$_3$:Pr (0.05%) crystal (Ø 2 mm, $d$ = 1 mm).

All procedures consisting of handling and weighing of all chemicals and manufacturing of grown crystals were performed in an atmosphere-controlled glovebox (MBraun LABstar unit) with the content of O$_2$ and H$_2$O below 1 ppm.

**Characterizations.** Powder X-ray diffraction (XRPD) was performed using a Rigaku MiniFlex 600 (Ni-filtered Cu K$\alpha_{1,2}$ radiation) equipped with a NaI:Tl scintillation detector. The XRPD patterns were compared to the relevant records in the ICDD PDF-2 database (version 2013). The angular range was 10–80°, with a step of 0.02° and a scanning speed of 2°/min. XRPD was measured to determine the phase purity of both the starting materials and the powdered parts of the grown crystals.

Absorption spectra were measured at room temperature (RT) using an ultraviolet/visible/near-infrared (UV/vis/NIR) spectrophotometer Shimadzu 3101PC in the range of 340–800 nm. Radioluminescence (RL), photoluminescence excitation (PLE), and photoluminescence emission (PL) spectra as well as PL decay curves were measured at RT and 77 K by a custom-made spectrofluorometer 5000M (Horiba Jobin Yvon, Wildwood, MA, USA) using the Mo X-ray tube (40 kV, 15 mA, Seifert), a steady-state xenon lamp (EQ-99X LDLS, Energetiq, a Hamamatsu Company), and nanosecond nanoLED pulsed light sources (fast prompt decay curves measured by the time-correlated single-photon counting technique at 5000M) as the excitation sources, respectively. The detection part of the setup involved a single-grating monochromator and a TBX-04 photon counting detector (IBH Scotland). Measured spectra were corrected for the spectral dependence of detection sensitivity (RL and PL) and the excitation light spectral dependence (PLE). The convolution procedure was applied to the photoluminescence decay curves to determine true decay times (SpectraSolve software package, Ames Photonics). The RL spectrum of the BGO (Bi$_4$Ge$_3$O$_{12}$) reference crystal scintillator was measured under identical geometrical conditions to obtain quantitative information on RL intensity of the samples.

The scintillation light yield is given by

$$LY = \frac{PhY}{\eta_{QE}\eta_{CE}} \quad (1)$$

where PhY is the photoelectron yield, i.e., the number of collected photoelectrons per MeV of absorbed energy, $\eta_{QE}$ is the quantum efficiency of a photodetector, and $\eta_{CE}$ is the light collection efficiency. Unfortunately, $\eta_{CE}$ cannot be determined for the sample studied due to the high absorption and nonhomogeneity. Therefore, we present the light yield estimate LY$_e$:

$$LY_e = \frac{PhY}{\eta_{QE}} \leq LY \quad (2)$$

The scintillation light yield estimate LY$_e$ was determined by pulse height spectroscopy of the scintillation response,[19] using a hybrid photomultiplier (HPMT),[20] model DEP PPO 475C; an ORTEC spectroscopy amplifier, model 672 (shaping time $t$ = 3 and 10 μs); and an ORTEC 927TM multichannel buffer. γ-Rays from $^{137}$Cs (662 keV) and α-particles from $^{239}$Pu (5.157 MeV) were used for excitation. The sample was optically coupled to the HPMT using silicon grease; more than 10 layers of Teflon tape were used as a reflector for experiments with $^{137}$Cs. α-Particle excitation did not enable reflector application. All measurements were performed at room temperature.

Samples were examined using a stereomicroscope, an Olympus SZX12 (100× magnification) equipped with a light source (operating in transmittance or reflectance mode), which was coupled with a Canon EOS5D digital camera for image capture. The camera was connected to a PC allowing live observation, image acquisition, and processing using QuickPHOTO software v 3.2.

Raman spectra were acquired on a Renishaw inVia confocal spectrometer, which operates inside a nitrogen-filled glovebox. For the measurements, a 633 nm Renishaw RL633 HeNe laser and a Leica 100× objective with numerical aperture NA = 0.90 and 1800 grooves/mm diffraction grating were used.

## RESULTS AND DISCUSSION

**X-ray Diffraction.** XRPD showed that the starting Lu$_2$S$_3$:Pr powders are phase-pure Lu$_2$S$_3$ (see Figure 2). No reflections belonging to the Lu$_2$O$_3$, Pr$_6$O$_{11}$, or Lu$_2$O$_2$S potential impurity phases were detected. The XRPD results of all grown Lu$_2$S$_3$:Pr crystals showed that Lu$_2$S$_3$ crystallizes in the rhombohedral ε-Ln$_2$S$_3$-type crystal structure (space group no. 167, phase prototype α-Al$_2$O$_3$), in accordance with the literature,[21] see Figure 3. Several minor reflections (strongest at $2\theta$ = 31.07°) in diffraction patterns of grown crystals can be ascribed to the Lu$_2$O$_2$S impurity phase. Since Lu$_2$O$_2$S is not present in the starting materials (see Figure 2), it is a result of a reaction of Lu$_2$S$_3$ with residual oxygen in the growth chamber at elevated temperatures.

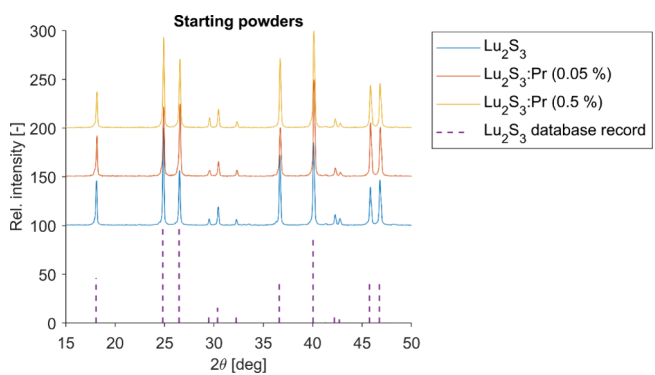

**Figure 2.** X-ray powder diffraction of Lu$_2$S$_3$:Pr ($x$ %) starting powders.

**Optical Properties.** Figure 4 shows the absorption spectra (at RT) of undoped and Pr$^{3+}$-doped (0.05%) Lu$_2$S$_3$ single crystals. The onset of the band-to-band transitions is well-visible at approximately 360 nm, between the values of 390 nm published by Schevciw and White[22] and of 330 nm published by van't Spijker et al.[15] Therefore, it is even low energy shifted with respect to the ternary sulfide KLuS$_2$, a band gap of which is located at around 303 nm.[23] The absorbance in the whole visible spectral range reaches high values due to scattering at imperfections (impurities and inclusions) in the prepared crystals, see Figure 1. Neither the characteristic 4f–5d transitions expected in the UV part of the spectra nor 4f–4f ones of the Pr$^{3+}$ ions are visible in the spectrum (Figure 4).

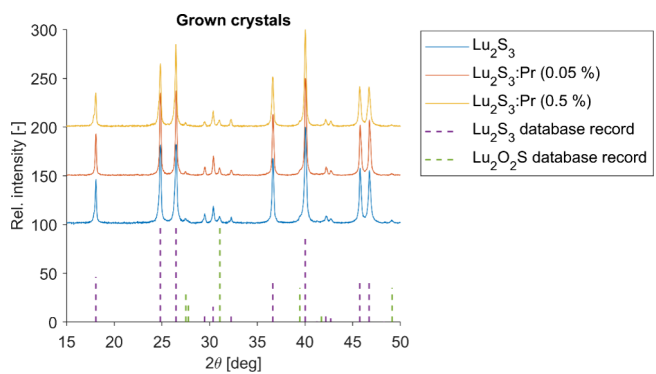

**Figure 3.** X-ray powder diffraction of the grown Lu$_2$S$_3$:Pr (x %) crystals. The value in parentheses corresponds to the nominal concentration of Pr in the melt.

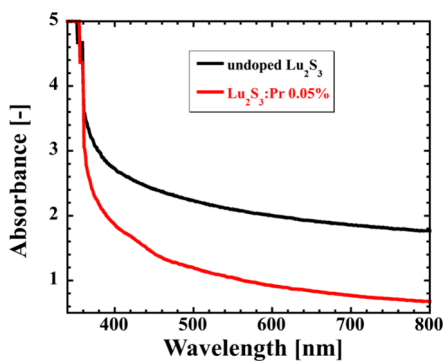

**Figure 4.** Absorption spectra of undoped (black solid line) and Pr$^{3+}$-doped (0.05%) (red solid line) Lu$_2$S$_3$ single crystals.

RT RL spectra of all three Lu$_2$S$_3$:Pr crystals are depicted in Figure S1. The undoped sample exhibits two broad bands peaking at 350 and 565 nm. Upon Pr$^{3+}$ doping, the host emission is suppressed, and the emission of 5d→4f and 4f→4f transitions of Pr$^{3+}$ appears. For the 0.05% sample, an overlap of host emission with Pr$^{3+}$ 4f→4f lines is clearly observed, while for the 0.5% doped sample, the contribution of host emission is either completely absent or too weak to be clearly distinguished. Therefore, Lu$_2$S$_3$:Pr$^{3+}$ (0.5%) was selected for further optical characterization. The RT RL spectrum of the Lu$_2$S$_3$:Pr 0.5% single crystal is shown in Figure 5a, together with the Bi$_4$Ge$_3$O$_{12}$ (BGO) single crystal scintillator serving as a reference material. Two different emissions can be recognized in the former spectrum. At the higher energy side, the Pr$^{3+}$ 5d→4f transitions can be assigned to the bands peaking between 360 and 445 nm. At RT, particular transitions cannot be well-resolved probably due to severe temperature quenching, see also below. In the lower energy part of the spectrum, between 480 and 800 nm, the partially relaxed parity-forbidden Pr$^{3+}$ 4f→4f transitions are well-visible. The RL spectrum integral for Lu$_2$S$_3$:Pr 0.5% (integrated in the 360–445 nm spectral region, where the fast 5d→4f emission is peaking) compared to that of the BGO standard scintillator (integrated in the 310–800 nm spectral region) shows that Lu$_2$S$_3$:Pr 0.5% provides only 8% of the BGO intensity at RT. The RL spectra of samples prepared from the start and middle of the Lu$_2$S$_3$:Pr (0.5%) crystal are very similar. The intensity of the Pr$^{3+}$ 5d→4f emission is unchanged, which suggests that the effective segregation coefficient of Pr in Lu$_2$S$_3$ is close to unity. However, detailed measurement of the Pr concentration profile along the crystal axis would be necessary to properly examine the segregation of Pr in Lu$_2$S$_3$. On the other hand, the relative intensity among 4f→4f lines is different in both samples. This could be caused by a decrease in transmittance in the green-red region or due to distortion of the Pr$^{3+}$ local environment resulting from an increasing concentration of defects toward the end of the crystal.

In contrast, at 77 K (see Figure 5b), emission bands in RL spectra corresponding to the 5d→4f emission of Pr$^{3+}$ become much better resolved and start to dominate over the Pr$^{3+}$ 4f–4f transitions. Therefore, the Pr$^{3+}$ 4f$^1$5d$^1$→$^3$H$_4$, 4f$^1$5d$^1$→$^3$H$_5$, and 4f$^1$5d$^1$→$^3$F$_{2,3}$ transitions can be assigned to the bands at 383, 413, and 464 nm, respectively, see Figure 5b. The significant increase in the intensity of 5d→4f emission with a decreasing temperature points to a heavy thermal quenching. Similar behavior was observed in Pr$^{3+}$-doped KLuS$_2$.[24,25] More pronounced temperature quenching of Pr$^{3+}$ 5d→4f emission in Lu$_2$S$_3$ compared to KLuS$_2$ is most probably the difference in the band gap width. In both compounds, the 5d orbitals of Lu predominantly contribute to the density of states at the bottom of the conduction band.[26] Therefore, "dilution" of the Lu sublattice by potassium in KLuS$_2$ should result in a blueshift of the bottom of the conduction band. This in turn results in higher energy separation between 5d levels of Pr$^{3+}$ and the bottom of the conduction band and thus lower temperature quenching. This corroborates with PLE spectra of intrinsic luminescence in undoped Lu$_2$S$_3$ and KLuS$_2$ (see Figure S2). The high energy offset of PLE for KLuS$_2$ is shifted toward shorter wavelengths compared to Lu$_2$S$_3$.

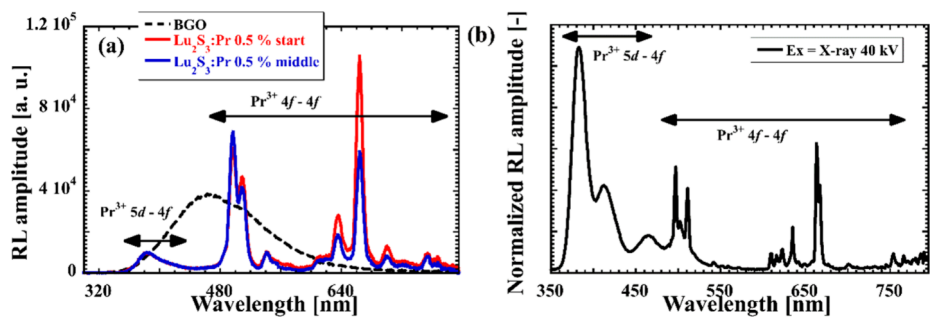

**Figure 5.** RL spectrum of Pr$^{3+}$-doped (0.5%) Lu$_2$S$_3$ single crystals recorded at (a) RT and (b) 77 K.

Figure 6 shows the PLE (for the 410 nm emission wavelength) and PL (under the 320 nm excitation wavelength)

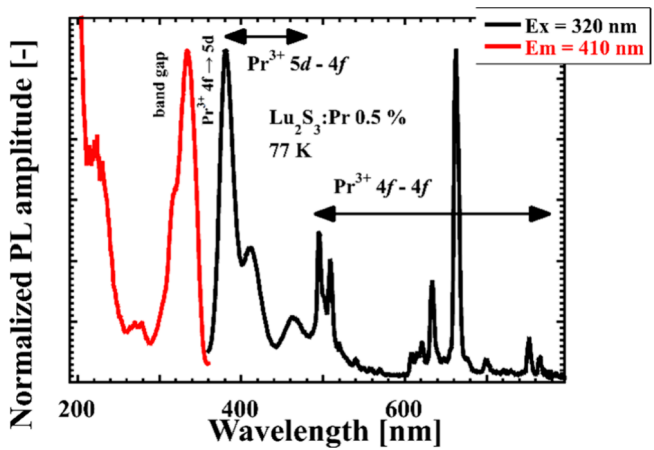

**Figure 6.** Photoluminescence excitation ($\lambda_{em}$ = 410 nm) and photoluminescence emission ($\lambda_{ex}$ = 320 nm) spectra of a $Pr^{3+}$-doped (0.5%) $Lu_2S_3$ single crystal recorded at 77 K.

spectra of the $Pr^{3+}$-doped (0.5%) $Lu_2S_3$ single crystal recorded at 77 K. The photoluminescence emission spectrum corresponds well with the spectrum excited by the X-ray, see Figure 5b, therefore showing both 5d→4f $Pr^{3+}$ transitions at 360−445 nm and 4f→4f $Pr^{3+}$ transitions in the 480−800 nm spectral region. The PLE spectrum shows a composed broad band peaking at 334 nm. The high energy part comprises a band gap (a high energy shift with respect to PLE at RT), while the lower energy part is given by the $Pr^{3+}$ 4f→5d absorption. PL decay curves corresponding to the 5d→4f transition ($\lambda_{ex}$ = 339 nm, $\lambda_{em}$ = 390 nm) recorded at 77 K and RT are shown in Figure 7a,b. The leading component of the decay curve at 77 K is characterized by the decay time of 10 ns, which is rather fast for the $Pr^{3+}$ 5d→4f transition, but an even shorter value of 7.5 ns was published for the $YAlO_3$ host.[27] At RT, see Figure 7b, the decay curve becomes nonexponential and significantly accelerated, with the leading component below 1 ns (∼0.2 ns). While such quenching might be deemed as negative, if the quenched $Pr^{3+}$ 5d→4f emission would be efficiently excited by ionizing radiation, it could open a way for application of $Lu_2S_3$:Pr as an ultrafast scintillator. Such materials are being searched for in connection with time-of-flight applications, such as positron emission tomography (TOF-PET) or computed tomography (TOF-CT).[28,29] Even if it gives only around 10% of BGO intensity under X-ray excitation (see Figure 4a), it is even faster than the recently introduced $KLuS_2$:Pr.[25] However, further study including measurement of scintillation decay kinetics and higher quality crystals with an optimized Pr concentration would be required to asses the potential of $Lu_2S_3$:Pr as an ultrafast scintillator.

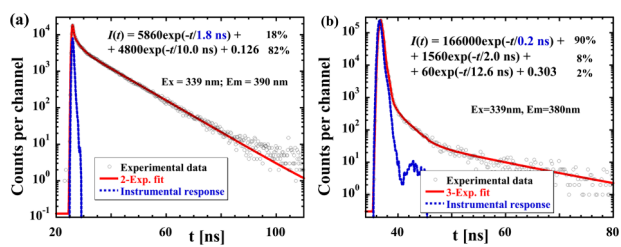

**Figure 7.** Photoluminescence decay ($\lambda_{ex}$ = 339 nm, $\lambda_{em}$ = 390 nm) of $Pr^{3+}$-doped (0.5%) $Lu_2S_3$ single crystals recorded at (a) 77 K and (b) RT.

Light yield measurement by an amplitude spectrometry setup was rather complicated. No peak is present in the amplitude spectrum of pulses produced by $^{137}Cs$ excitation, only continuum. Nevertheless, the amplitude spectrum can be surely ascribed to the $^{137}Cs$ excitation, see spectra of the sample measured with and without the $^{137}Cs$ source. The $^{137}Cs$ excited spectrum is distorted due to the nonhomogeneous nature of the sample and strong overlap of absorption and emission spectra, see Figures 4 and 5. Therefore, the scintillation photon collection efficiency depends on the depth of interaction of γ photons inside the sample. If the interaction occurs closer to HPMT photocathode, the scintillation photon collection efficiency is higher and vice versa. Since 662 keV photons are only slightly absorbed by the sample, all interactions depths are of approximately the same probability. Thus, only the light yield lower bound can be estimated, $LY_e$ (γ 662 keV) ≥ 2300 ph/MeV, using the end point value in amplitude spectra, see Figure 8.

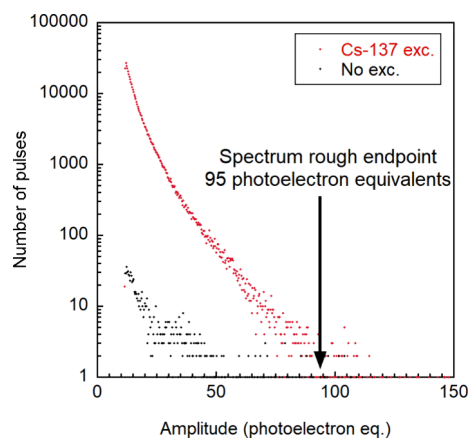

**Figure 8.** Amplitude spectra of sample $Lu_2S_3$:Pr (0.5%) measured by HPMT, excited by $^{137}Cs$ and without any excitation.

In the α-particle-excited amplitude spectra, a highly deformed peak is observable on the other hand, see Figure S3. α-Particles deposit their entire energy near the sample surface (the side far from the HPMT photocathode). Therefore, variations in the scintillation photon collection efficiency are significantly lower, and the amplitude spectrum is less distorted. From the position of the photopeak, the lower bound of light yield can be estimated to be $LY_e(\alpha)$ ≥ 190 ph/MeV. Please note that $LY_e(\alpha)$ is affected by the fact that no reflector was used, which may decrease the scintillation photon collection efficiency very roughly by an additional 30%.[30] Also, scintillation photons produced by an α-particle excitation are absorbed more since they must pass the entire thickness of the sample.

The above presented $LY_e$ values are for the 3 μs shaping time. The value measured using the 10 μs shaping time differs by 5% maximally, which suggests that most of the scintillation photons were able to contribute to the signal.

The $Pr^{3+}$-doped $Lu_2S_3$ (0.05%) single crystal was characterized by Raman spectroscopy; the resulting spectra excited by the 633 nm laser are depicted in Figure 9 in the energy range of 55−500 cm$^{-1}$. As shown by Rademaker,[31] the highest observed energy peak can be a first approximation for effective phonon energy of host crystalline matrices. For the $Pr^{3+}$-doped $Lu_2S_3$ (0.05%) single crystal, it was therefore estimated to be an $E_{ph}$ of ∼312 cm$^{-1}$, which is comparable to other low-phonon materials like ZnSe ($E_{ph}$ of ∼250 cm$^{-1}$),[32] $KLuS_2$ ($E_{ph}$ of ∼220 cm$^{-1}$),[33] or $KPb_2Cl_5$ ($E_{ph}$ of ∼200 cm$^{-1}$).[34] This reduces the multiphonon quenching and should minimize the nonradiative losses of lasing ions especially in the mid-infrared spectral region.

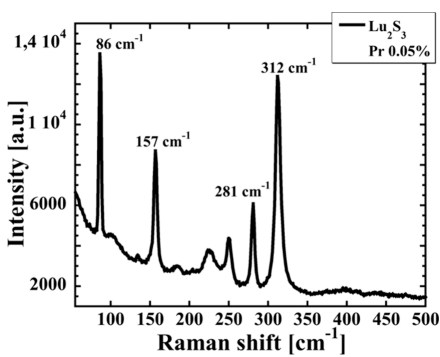

**Figure 9.** Raman spectra of $Pr^{3+}$-doped $Lu_2S_3$ (0.05%) excited by a 633 nm laser.

## ■ CONCLUSIONS

We present the first report on the successful growth of $Lu_2S_3$ (undoped and Pr-doped) single crystals by the micropulling-down method. To the best of our knowledge, this is the first report of $Lu_2S_3$ crystal growth from a melt. To further demonstrate the practical capabilities of grown crystals, we have selected Pr doping as a proof-of-concept system for luminescence and scintillation studies. It was shown that $Lu_2S_3$:Pr exhibits both fast 5d→4f and slow 4f→4f radiative transitions of the $Pr^{3+}$ center. Even though the 5d→4f transitions of $Pr^{3+}$ are heavily quenched at RT in $Lu_2S_3$, it is worth studying the characteristics of other RE dopants showing 5d→4f emissions as the $Ce^{3+}$, $Eu^{2+}$, and $Sm^{2+}$. The technology presented above opens a window for exploration of a large material group ({RE}$_2S_3$ and other high melting sulfides) in a bulk single-crystal form, some of which could be interesting candidates for applications in scintillation detectors or as laser active media.

## ■ ASSOCIATED CONTENT

### ⓈSupporting Information

The Supporting Information is available free of charge at https://pubs.acs.org/doi/10.1021/acs.cgd.4c00330.

　　Radioluminescence spectra, photoluminescence excitation spectra of undoped samples, and amplitude spectra of sample $Lu_2S_3$:Pr (0.5%) excited by α-particles of $^{239}$Pu (PDF)

## ■ AUTHOR INFORMATION

### Corresponding Author

**Vojtěch Vaněček** − Institute of Physics of the Czech Academy of Sciences, Prague 6 162 00, Czech Republic; Faculty of Nuclear Sciences and Physical Engineering, Czech Technical University in Prague, Prague 1 115 19, Czech Republic; orcid.org/0000-0001-9730-9570; Email: vanecekv@fzu.cz

### Authors

**Vítězslav Jarý** − Institute of Physics of the Czech Academy of Sciences, Prague 6 162 00, Czech Republic; orcid.org/0000-0002-5149-7307

**Robert Král** − Institute of Physics of the Czech Academy of Sciences, Prague 6 162 00, Czech Republic; orcid.org/0000-0002-4519-6030

**Lubomír Havlák** − Institute of Physics of the Czech Academy of Sciences, Prague 6 162 00, Czech Republic; orcid.org/0000-0002-0160-4522

**Aleš Vlk** − Institute of Physics of the Czech Academy of Sciences, Prague 6 162 00, Czech Republic; orcid.org/0000-0003-2866-7133

**Romana Kučerková** − Institute of Physics of the Czech Academy of Sciences, Prague 6 162 00, Czech Republic; orcid.org/0000-0001-9441-0681

**Petr Průša** − Institute of Physics of the Czech Academy of Sciences, Prague 6 162 00, Czech Republic; Faculty of Nuclear Sciences and Physical Engineering, Czech Technical University in Prague, Prague 1 115 19, Czech Republic; orcid.org/0000-0002-2766-2907

**Jan Bárta** − Institute of Physics of the Czech Academy of Sciences, Prague 6 162 00, Czech Republic; Faculty of Nuclear Sciences and Physical Engineering, Czech Technical University in Prague, Prague 1 115 19, Czech Republic; orcid.org/0000-0001-9745-5367

**Martin Nikl** − Institute of Physics of the Czech Academy of Sciences, Prague 6 162 00, Czech Republic; orcid.org/0000-0002-2378-208X

Complete contact information is available at:
https://pubs.acs.org/10.1021/acs.cgd.4c00330

### Notes

The authors declare no competing financial interest.

## ■ ACKNOWLEDGMENTS

The work is supported by Operational Programme Johannes Amos Comenius financed by European Structural and Investment Funds and the Czech Ministry of Education, Youth, and Sports (Project No. SENDISO-CZ.02.01.01/00/22_008/0004596). We acknowledge the use of the CzechNanoLab research infrastructure (LM2023051) supported by the MEYS.

## ■ REFERENCES

(1) Kamarzin, A. A.; Mironov, K. E.; Sokolov, V. V.; Malovitsky, Yu. N.; Vasil'yeva, I. G. Growth and Properties of Lantanum and Rare-Earth Metal Sesquisulfide Crystals. *J. Cryst. Growth* **1981**, *52*, 619−622.

(2) Range, K.-J.; Lange, K. G.; Drexler, H. Structural Relations and Phase Transformations in the Rare-Earth Sesquisulfide Series at High Pressures and Temperatures. *Comments on Inorganic Chemistry* **1984**, *3* (4), 171−188.

(3) Kanazawa, M.; Li, L.; Kuzuya, T.; Takeda, K.; Hirai, S.; Higo, Y.; Shinmei, T.; Irifune, T.; Sekine, C. High-Pressure and High-Temperature Synthesis of Heavy Lanthanide Sesquisulfides $Ln_2S_3$ (Ln = Yb and Lu). *J. Alloys Compd.* **2018**, *736*, 314−321.

(4) Andreev, P. O.; Polkovnikov, A. A.; Denisenko, Y. G.; Andreev, O. V.; Burkhanova, T. M.; Bobylevb, A. N.; Pimneva, L. A. Temperatures and Enthalpies of Melting of $Ln_2S_3$ (Ln = Gd, Tb, Dy, Ho, Er, Tm, Yb, and Lu) Compounds. *J. Therm Anal Calorim* **2018**, *131* (2), 1545−1551.

(5) Besançon, P. Teneur En Oxygéne et Formule Exacte d'une Famille de Composés Habituellement Appelés "Variété β" Ou "Phase

Complexe" Des Sulfures de Terres Rares. *J. Solid State Chem.* **1973**, 7 (2), 232−240.

(6) Fang, C. M.; Meetsma, A.; Wiegers, G. A.; Boom, G. Synthesis and Crystal Structure of F-Type Erbium Sesquisulfide, F-ER2S3. *J. Alloys Compd.* **1993**, 201 (1), 255−259.

(7) Kamarzin, A.; Sokolov, V.; Mironov, K. Physicochemical Properties of StoichiometricLa2S3 Single Crystals. *Mater. Res. Bull.* **1976**, 11, 695−698.

(8) Henderson, J. R.; Muramoto, M.; Loh, E.; Gruber, J. B. Electronic Structure of Rare-Earth Sesquisulfide Crystals. *J. Chem. Phys.* **2004**, 47 (9), 3347−3356.

(9) Leiss, M. Nd3+ Photoluminescence in Semiconducting Binary Sesquisulphide Crystals of La, Gd and Y. *J. Phys. C: Solid State Phys.* **1980**, 13 (1), 151.

(10) Prewitt, C. T.; Sleight, A. W. Structure of Gadolinium Sesquisulfide. *Inorg. Chem.* **1968**, 7 (6), 1090−1093.

(11) White, J. G.; Yocom, P. N.; Lerner, S. Structure Determination and Crystal Preparation of Monoclinic Rare Earth Sesquisulfides. *Inorg. Chem.* **1967**, 6 (10), 1872−1875.

(12) Ebisu, S.; Era, T.; Guo, Q.; Miyazaki, M. Extremely Anisotropic Suppression of Huge Enhancement of Electrical Resistivity by Magnetic Field in α-R2S3 (R = Sm, Dy). *J. Phys.: Conf. Ser.* **2018**, 969 (1), No. 012124.

(13) Ebisu, S.; Iijima, Y.; Iwasa, T.; Nagata, S. Antiferromagnetic Transition and Electrical Conductivity in α-Gd2S3. *J. Phys. Chem. Solids* **2004**, 65 (6), 1113−1120.

(14) Range, K. J.; Leeb, R. HHigh Pressure Modifications of the Rare Earth Sulfides Ln$_2$S$_3$ (Ln = Lu-Ho, Y) with U$_2$S$_3$-Structure High Pressure Modifications of the Rare Earth Sulfides Ln$_2$S$_3$ (Ln = Lu-Ho, Y) with U$_2$S$_3$-Structure. *Z. Naturforsch. B* **1975**, 30 (6), 889−895.

(15) van't Spijker, J. C.; Dorenbos, P.; Allier, C. P.; van Eijk, C. W. E.; Ettema, A. R. H. F.; Huber, G. Lu2S3:Ce3+, A New Red Luminescing Scintillator. *Nuclear Instruments and Methods in Physics Research Section B: Beam Interactions with Materials and Atoms* **1998**, 134 (2), 304−309.

(16) Cunningham, G.; Shen, Y.; Bray, K. L. Effect of Pressure on Luminescence Properties of Ce3+:Lu2S3. *Phys. Rev. B* **2001**, 65 (2), No. 024112.

(17) Kral, R.; Jary, V.; Pejchal, J.; Kurosawa, S.; Nitsch, K.; Yokota, Y.; Nikl, M.; Yoshikawa, A. Growth and Luminescence Properties of Single Crystals Prepared by Modified Micro-Pulling-Down Method. *IEEE Trans. Nucl. Sci.* **2016**, 63 (2), 453−458.

(18) *Shaped Crystals: Growth by Micro-Pulling-Down Technique*; Fukuda, T.; Chani, V. I., Eds.; Kawazoe, Y., Hasegawa, M., Inoue, A., Kobayashi, N., Sakurai, T., Wille, L., Series Eds.; Advances in Materials Research; Springer: Berlin, Heidelberg, 2007; Vol. 8. DOI: 10.1007/978-3-540-71295-4.

(19) Moszynski, M.; Kapusta, M.; Mayhugh, M.; Wolski, D.; Flyckt, S. O. Absolute Light Output of Scintillators. *IEEE Trans. Nucl. Sci.* **1997**, 44 (3), 1052−1061.

(20) D'Ambrosio, C.; Leutz, H. Hybrid Photon Detectors. *Nuclear Instruments and Methods in Physics Research Section A: Accelerators, Spectrometers, Detectors and Associated Equipment* **2003**, 501 (2), 463−498.

(21) Ferrand, B.; Grange, Y. Apparatus for Making a Single Crystal. EP0130865A1, January 9, 1985.

(22) Schevciw, O.; White, W. B. The Optical Absorption Edge of Rare Earth Sesquisulfides and Alkaline Earth - Rare Earth Sulfides. *Mater. Res. Bull.* **1983**, 18 (9), 1059−1068.

(23) Jarý, V.; Havlák, L.; Bárta, J.; Mihóková, E.; Buryi, M.; Nikl, M. ALnS2:RE (A = K, Rb; Ln = La, Gd, Lu, Y): New Optical Materials Family. *J. Lumin.* **2016**, 170, 718−735.

(24) Havlák, L.; Jarý, V.; Rejman, M.; Mihóková, E.; Bárta, J.; Nikl, M. Luminescence Characteristics of Doubly Doped KLuS2:Eu, RE (RE = Pr, Sm, Ce). *Opt. Mater.* **2015**, 41, 94−97.

(25) Jarý, V.; Havlák, L.; Bárta, J.; Mihóková, E.; Kučerková, R.; Buryi, M.; Babin, V.; Průša, P.; Vrba, T.; Kotlov, A.; Nikl, M. Efficient Ultrafast Scintillation of KLuS 2: Pr 3 + Phosphor: A Candidate for Fast-Timing Applications. *Phys. Rev. Appl.* **2023**, 19 (3), No. 034092.

(26) He, L.; Meng, J.; Feng, J.; Zhang, Z.; Liu, X.; Zhang, H. Insight into the Characteristics of 4f-Related Electronic Transitions for Rare-Earth-Doped KLuS2 Luminescent Materials through First-Principles Calculation. *J. Phys. Chem. C* **2020**, 124 (1), 932−938.

(27) Zych, A.; de Lange, M.; de Mello Donegá, C.; Meijerink, A. Analysis of the Radiative Lifetime of Pr3+ D-f Emission. *J. Appl. Phys.* **2012**, 112 (1), No. 013536.

(28) Rossignol, J.; Turtos, R. M.; Gundacker, S.; Gaudreault, D.; Auffray, E.; Lecoq, P.; Bérubé-Lauzière, Y.; Fontaine, R. Time-of-Flight Computed Tomography - Proof of Principle. *Phys. Med. Biol.* **2020**, 65 (8), No. 085013.

(29) Lecoq, P.; Morel, C.; Prior, J. O.; Visvikis, D.; Gundacker, S.; Auffray, E.; Križan, P.; Turtos, R. M.; Thers, D.; Charbon, E.; Varela, J.; de La Taille, C.; Rivetti, A.; Breton, D.; Pratte, J.-F.; Nuyts, J.; Surti, S.; Vandenberghe, S.; Marsden, P.; Parodi, K.; Benlloch, J. M.; Benoit, M. Roadmap toward the 10 Ps Time-of-Flight PET Challenge. *Phys. Med. Biol.* **2020**, 65 (21), 21RM01.

(30) Prusa, P.; Kučera, M.; Babin, V.; Bruza, P.; Parkman, T.; Panek, D.; Beitlerova, A.; Mares, J. A.; Hanus, M.; Lucenicova, Z.; Pokorny, M.; Nikl, M. Tailoring and Optimization of LuAG:Ce Epitaxial Film Scintillation Properties by Mg Co-Doping. *Cryst. Growth Des.* **2018**, 18 (9), 4998−5007.

(31) Rademaker, K. *Rare Earth-Doped Alkali-Lead-Halide Laser Crystals of Low-Phonon Energy*; Cuvillier Verlag, 2005.

(32) Taylor, W. The Raman Spectra of Cubic Zinc Selenide and Telluride. *Phys. Lett. A* **1967**, 24 (10), 556−558.

(33) Šulc, J.; Švejkar, R.; Fibrich, M.; Jelínková, H.; Havlák, L.; Jarý, V.; Ledinský, M.; Nikl, M.; Bárta, J.; Buryi, M.; Lorenzi, R.; Cova, F.; Vedda, A. Infrared Spectroscopic Properties of Low-Phonon Lanthanide-Doped KLuS2 Crystals. *J. Lumin.* **2019**, 211, 100−107.

(34) Tkachuk, A. M.; Ivanova, S. É.; Isaenko, L. I.; Yelisseyev, A. P.; Joubert, M.-F.; Guyot, Y.; Payne, S. Spectroscopic Studies of Erbium-Doped Potassium-Lead Double Chloride Crystals KPb2Cl5:Er3+: 1. Optical Spectra and Relaxation of Excited States of the Erbium Ion in Potassium-Lead Double Chloride Crystals. *Opt. Spectrosc.* **2003**, 95 (5), 722−740.